# Influences of social media usage on public attitudes and behavior toward COVID-19 vaccine in the Arab world

Md. Rafiul Biswas, Hazrat Ali, Raian Ali & Zubair Shah

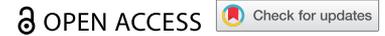

RESEARCH PAPER　　　　　　　　　　　　　　　　　　　　　　　　　　　　OPEN ACCESS

# Influences of social media usage on public attitudes and behavior toward COVID-19 vaccine in the Arab world

Md. Rafiul Biswas, Hazrat Ali, Raian Ali, and Zubair Shah

Division of Information and Computing Technology, College of Science and Engineering, Hamad Bin Khalifa University, Qatar Foundation, Doha, Qatar

**ABSTRACT**
**Background:** Vaccination programs are effective only when a significant percentage of people are vaccinated. Social media usage is arguably one of the factors affecting public attitudes toward vaccines.
**Objective:** This study aims to identify if the social media usage factors can predict Arab people's attitudes and behavior toward the COVID-19 vaccines.
**Methods:** An online survey was conducted in the Arab countries, and 217 Arab nationals participated in this study. Logistic regression was applied to identify what demographics and social media usage factors predict public attitudes and behavior toward the COVID-19 vaccines.
**Results:** Of the 217 participants, 56.2% (n = 122) were willing to get the vaccines, and 41.5% (n = 90) were hesitant. This study shows that none of the social media usage factors were significant enough to predict the actual vaccine acceptance behavior. However, some social media usage factors could predict public attitudes toward the COVID-19 vaccines. For example, compared to infrequent social media users, frequent social media users were 2.85 times more likely to agree that the risk of COVID-19 was being exaggerated (OR = 2.85, 95% CI = 0.86–9.45, $p$ = .046). On the other hand, participants with more trust in vaccine information shared by their contacts were less likely to agree that decision-makers had ensured the safety of vaccines (OR = 0.528, 95% CI = 0.276–1.012, $p$ = .05).
**Conclusion:** Information shared on social media may affect public attitudes toward COVID-19 vaccines. Therefore, disseminating correct and validated information about the COVID-19 vaccines on social media is important to increase public trust and counter the impact of incorrect misinformation.

## Introduction

The World Health Organization (WHO) continuously monitors the situation of COVID-19 and provides guidelines to contain the virus. To mitigate the spread of coronavirus, preventive measures, such as prohibiting social gatherings, social distancing, wearing a mask, and travel bans were enforced in the majority of the countries.[1] Vaccination is an effective option to confront the virus and overcome the crisis. By the end of 2021, WHO has approved eight vaccines, namely Modern, Pfizer/BioNTech, Janssen (Johnson & Johnson), Oxford/AstraZeneca, Covishield, Covaxin, Sinopharm, and Sinovac.[2] These vaccines have proven effective and safe, but only a high level of vaccination can reduce the spread of the virus in the population and avoid disruptive effects on health systems. However, the COVID-19 vaccine introduction has been accompanied by a high level of vaccine hesitancy. The vaccine hesitancy delays the vaccination despite the availability of the vaccine and recommendations of the authorities.[3] WHO declared vaccine hesitancy among the top ten major threats against infectious disease.[4] Therefore, it is important to measure vaccine acceptance and hesitancy rate and the underlying attitudes toward vaccination. Vaccine hesitancy is linked to several factors, such as external factors (i.e., immunization requirements, social belief, policies, and media), vaccine-specific factors (i.e., vaccine efficacy, safety, and susceptibility), and host-specific factors (i.e., education level, knowledge about vaccine, and income).[3] Since 2020, several surveys have reported that vaccine acceptance rates have varied among different countries and ethnic groups. High rates of vaccine acceptance were reported in Ecuador (97%), Malaysia (94.3%), China (91.3%), whereas low rates of vaccine uptake were reported in Italy (53.7), Russia (54.9%), Poland (56.3%), and US (56.9%).[5,6] Vaccine acceptance was reported above 60% in three Arab countries, Saudi Arabia,[7] Qatar,[8] and UAE,[9] above 50% in Kuwait[10] and Oman,[11] and below 40% in Jordan.[12]

Social media played a significant role in disseminating health-related information during the COVID-19 pandemic. Every 45 million, at least one tweet related to COVID-19 was shared.[13] A comprehensive range of health information, including both evidence-informed health communications as well as misinformation and misinformed opinions, spread fast through social media.[14] Ongoing exposure to health information in the news and social media may also shape people's attitudes and behavior and affect vaccination decision-making.[15] In addition, different cultural beliefs and family traditions may also lead to unawareness and resistance.

Attitude toward vaccination is one key example of the challenges of addressing persistent misinformation in public health, but others continue to emerge. Studies suggest that negative information about vaccines from news media, health practitioners, and celebrities can increase vaccine refusal and hesitancy,[16] and affect vaccine uptake and vaccination coverage. The Strategic Advisory Group of Experts on Immunization (SAGE) reported that communication strongly affects vaccine acceptance.[17] SAGE also reported that poor communication undermines vaccine uptake.[17] Sharing health-related authentic information through social media can motivate people to be vaccinated against coronavirus. A huge number of studies have been conducted to find out the socioeconomic determinants of vaccine hesitancy. However, there have not been a good number of worldwide cross-sectional analyses to find the effect of social media in vaccine hesitancy.[18,19] Through an online survey in Arab countries, this study aims to verify if social media usage can predict respondents' intention or attitudes toward vaccination against COVID-19.

## Methods

### Study design and data collection

An online survey was conducted in the Arab region between 2 May 2021, and 25 June 2021, to determine the potential of social media usage to predict attitudes and behavior toward the COVID-19 vaccines. The Institutional Review Board at Qatar Biomedical Research Institute (QBRI), Hamad Bin Khalifa University, approved the study (Ref. QBRI-IRB 2021-03-083). Participants of the study were informed of the purpose of the study, and they were entirely free to take part or opt-out as they wished. No personal information like name, e-mail, or id were collected. However, voluntarily the name of the Twitter account of the respondent was collected. People from different Arab countries (e.g., Saudi Arabia, Qatar, Oman, Jordan, and Kuwait) participated in the survey. Participants were asked to complete an online survey form that included questions in both English and Arabic language.

This study was performed based on a convenience sampling methodology. The survey was shared as a link through different mailing lists and social media platforms (e.g., Twitter, Facebook, WhatsApp, LinkedIn, Instagram). Participants were encouraged to share the survey link with their friends, colleagues, and contacts. It is speculated that people might have shared the link within their circles, and snowballing could have happened. The survey was performed when vaccines were available in these countries. It was assumed that people had some knowledge about the COVID-19 vaccines. The survey link was shared through online media, and we don't know the number of people who received this link. Hence, the response rate was not calculated. We note that the survey link was primarily shared in the gulf area, so the responses mainly came from the GCC countries. Only people of age 18 years or above were considered for this study. Incomplete answers, quickly completed responses, and questions presenting an answers pattern were excluded from the study. A total of 317 respondents from different countries participated in the survey, of which only 217 fulfilled the study criteria. The mean time of survey completion was about 5 min.

### Questionnaire development

The questionnaire covered four main themes: a) demographics, ii) social media, iii) COVID-19 vaccine acceptance, and iv) attitudes toward COVID-19 vaccines. The demographics questions are standard questions and are commonly used in literature.[7,12,20] The vaccine acceptance questions were obtained from the KFF Health Tracking Poll study.[21] Vaccine-specific questions were widely used in literature[8–22–24] to determine the determinants of vaccine hesitancy. Social media-related questions were developed based on existing literature[19,25,26] and with the help of two expert researchers in the field. The survey was tested on a sample of five bilingual individuals before distributing it to actual respondents. We refined the language used in the survey according to their comments. The data were kept confidential and were only accessible to authors named in the IRB approval.

### Demographics questions

Demographic questions consisted of age, gender, nationality, country of residence, and living area. It also included questions related to occupations, education, and type of living. There was a free text field for the participants to provide input for age. There were three options for the gender question: male, female, and prefer not to say. A few Arab countries were listed in the options where the COVID-19 vaccines were made widely available, and there was a free text field to provide input for other countries not listed. Participants were asked to choose their living areas, urban or rural. They were also asked about their type of living, such as living with family, living alone, and prefer not to say. They were further asked to select their occupation from the list, such as students, health professionals, educationalists, employed in other sectors, own businesses, and unemployed. Appendix A provides the complete questionnaire form.

### Social media questions

There were five questions related to social media that aimed to elicit how people use social media and search for health-related information online. Participants were asked about their frequency of social media usage, how often they search vaccine-related information online and how much they trust vaccine-related information if their contact on social media share it. Participants were also asked how they reacted to negative comments about vaccines on social media and their level of engagement in vaccine-specific discussions on social media.

### COVID-19 vaccines acceptance questions

There were two questions to measure COVID-19 vaccine intention. Firstly, participants were asked about their willingness to vaccinate when the vaccines are recommended by health professionals and made available for their age group, profession, and health condition. Participants who respond as either very likely to get it or likely to get it are further asked another question to see if they will get it as soon as possible. Otherwise, they wanted to delay it to see if vaccines work for others or get the vaccine if required for their work, school, or travel.

We focused on whether the participants had been hesitant or willing to take the vaccine. Being vaccinated does not necessarily imply that the participant was willing to take the vaccine. Many individuals took the vaccine due to conformity and compliance with job and travel requirements and access to offices, public transportation, or schools.[27,28] We found that some of those respondents who took the COVID-19 vaccine were among the respondents who had various concerns about the COVID-19 vaccines.

### Vaccine-specific questions

There were 11 questions to measure the vaccine-specific attitudes of participants. The responses to these questions were measured on a five-point Likert scale from fully disagree to fully agree. The questions were about the perceived risk of COVID-19 infection, the efficacy of the COVID-19 vaccine, trust in the government and manufacturing of COVID-19 vaccines, and knowledge about the vaccine. The questions also investigated public concerns about vaccine safety and side effects. In addition, there was a free text box option to write views about the COVID-19 vaccines. These questions were compiled from several studies.[23,29,30]

## Data analysis

### Data preprocessing

*Data cleaning.* Data were preprocessed by eliminating missing values, removing duplicates, and excluding data that did not meet the criteria (e.g., responses completed in less than half of the mean completion time). Also, 317 respondents completed the survey, of which 220 were from 18 Arab nationalities, of which three participants had aged less than 18 years and were removed. Finally, 217 participants were selected for analysis. Both Arabic and English responses were considered for analysis. Responses in Arabic were translated to English, and the final data was encoded for analysis.

*Data encoding.* Responses to various questions were obtained in continuous, ordinal, and categorical formats. The responses were encoded into numbers to represent them as categorical variables. The responses to questions were encoded by organizing one or more responses into a specific category. For example, the age variable contains responses in continuous numerical value, which was classified into three age categories, such as 18–25 years, 26–35 years, and >35 years following a paper.[12] The responses to questions related to the frequency of social media usage and frequency of searching vaccine-related information online were classified into two categories frequent and nonfrequent. The responses to questions related to trust in vaccine-related information shared by their contacts were classified into two categories as trust and no trust. Similarly, the responses to the remaining two questions about social media were divided into two categories active and passive.

The responses to questions 14 and 15 were combined hierarchically to form a computed variable *COVID-19 vaccine intention* that has two categories: acceptant and hesitant. The acceptant group was computed as when the response to question 14 was either *very likely to get it* or *likely to get it*, and response to question 15 was either *get it as soon as possible* or *already got vaccine* indicated under *other* option. Similarly, the hesitant group was determined when one of the following two conditions was true. First, when the response to question 14 was any of the options, *very unlikely to get it* or *unlikely to get it or unsure*. Second, when the response to question 14 was either *very likely to get it* or *likely to get it* but also a response to question 15 was either *wait until it has been available for a while to see if it is working for other people* or *only get the vaccine if it is required to do so for work, school, or other activities*. All other combinations of responses, such as *don't know*, were not considered for any of the two groups, namely hesitant or acceptant.

Finally, the responses for the vaccine-specific questions were classified into two categories; disagree and agree for binary logistic regression and three categories as disagree, neutral, and agree for multinomial logistic regression. The detailed encoding scheme is shown in Appendix B.

### Statistical analysis

The frequency, mean, maximum, mode, and standard deviation were measured to find the descriptive characteristics of the data. Independent variables were considered from the seven questions related to age, gender, and five social media-related questions (e.g., frequency of social media usage and searching vaccines-related information online, etc.). Dependent variables were considered from one computed variable *COVID-19 vaccine intention* (calculated from questions 14 and 15) and 11 vaccine-specific questions. Before applying binary logistic regression, the assumption was checked if the independent variables met the assumption criteria (i.e., case predictor ratio, multicollinearity, outliers). To perform binary logistic regression, one dependent variable and one or more than one independent variable were used at a time. Binary logistic regression was applied to identify the factors from seven independent variables that affect COVID-19 vaccine intention or vaccine attitudes from 12 dependent variables.

Positive responses (i.e., agree, frequent, trust and active, etc.) were encoded as class one for the dependent variables. For the independent variables, negative responses (e.g., not frequent, no trust and passive, etc.) were used as a reference group. All reference variables were used as a base for comparison. In logistic regression, the results were reported with Odd Ratios (OR) and 95% confidence intervals (CI). The level of statistical significance was set to 5%. Also, multinomial logistic regression was applied where neutral was included with the two groups, positive and negative. In multinomial logistic regression, the level of agreement was compared with neutral. Also, multiple independent variables (i.e., maximum three) were combined to find factors that can predict vaccine intention and public attitudes toward COVID-19 vaccines. JASP statistical package JASP 0.15 was used to find descriptive statistics and perform binary logistic regression, whereas IBM SPSS statistical package v.25 was used for multinomial logistic regression.

## Results

A total of 317 participants completed the survey. 217 participants were from 18 Arab nationalities who met the study criteria. The following analysis is based on responses received from these participants.

## Demographics characteristics

Table 1 shows the demographic characteristics of the participants. Of the 217 participants, 60.4% (n = 131) were male, and the median age of participants was 34 years. 88% (n = 191) of the participants lived with their family and 89.4% (n = 194) lived in urban areas. 41.5% (n = 90) of the participants were employed in different sectors (e.g., private job, NGO job, freelancing, and so on), 17.1% (n = 37) were students, while the rest of them belonged to different professions, as summarized in Table 1. 75.1% (n = 163) of the participants were from Saudi Arabia, Qatar, Oman, and Jordan and the remaining of the participants were from the other 14 Arab countries (Algeria, Egypt, Iraq, Kuwait, Lebanon, Libya, Morocco, Palestine, Somalia, Sudan, Syria, Tunisia, United Arab Emirates, and Yemen).

89.4% (n = 194) of the participants were frequent on social media, whereas 14.8% (n = 32) of the participants were frequent in searching for vaccine information online. 58.9% (n = 128) of the participants trusted the COVID-19 vaccine information shared by their contacts. 63.6% (n = 138) of the participants actively reacted (e.g., ask health professionals, ask friends, search for information on the internet) when they heard any negative comments about the vaccine from social media. 17.5% (n = 38) of the participants actively participated (e.g., reply, comment, share) in the COVID-19 vaccine discussion on social media.

## Descriptive statistics for vaccine acceptance

Table 2 shows the detailed descriptive statistics of vaccine acceptance responses of the participants. 217 participants answered question 14. Based on a computed variable, *COVID-19 vaccine intention*, derived from questions 14 and 15, 56.2% (n = 122) of the participants were vaccine-acceptant, and 41.8% (n = 90) of them were vaccine-hesitant. Although 86.7% (n = 188) of the participants responded to question 14 that *they were likely to get the vaccine* but 32.4% (n = 61) of those 188 respondents mentioned in question 15 that *they would get the vaccine only if required for their work or wait to see if it worked for others*. Of 188 participants, 64.9% (n = 122) were in the vaccine-acceptant group, 32.44% (n = 61) were in the vaccine-hesitant group, and the remaining 2.65% (n = 5) responded as *don't know*. We did not categorize the five participants ('don't know category') into vaccine acceptance or the vaccine-hesitant group.

## Social media and COVID-19 vaccine acceptance

Binary logistic regression was applied to predict social media usage on the likelihood that social media usage might affect vaccine acceptance, see Table 3. The assumptions were checked before applying binary logistic regression. Firstly, the sample size was more significant than the predictor size. Secondly, there was weak multicollinearity (r = 0.219) between two independent variables *hearing negative comments about vaccines* and *engaging with vaccine discussion on social media*. So, these two independent variables were not selected together to avoid multicollinearity. Thirdly, there were no outliers. Thus, the independent variables met the assumption criteria. The result shows that none of the social media usage factors are significant enough (i.e., $p$-value <.05) to predict the vaccine acceptance behavior of the participants.

Table 1. Demographic characteristics of participants (n = 217).

| Characteristic variable | Frequency (Percentage) |
|---|---|
| Age | |
|   Median | 34 |
|   18–25 years | 48 (22.1) |
|   26–35 years | 77 (35.5) |
|   >35 years | 92 (42.4) |
| Gender | |
|   Male | 131 (60.4) |
|   Female | 84 (38.7) |
|   Prefer not to say | 2 (0.9) |
| Live with | |
|   Alone | 20 (9.2) |
|   Family | 191 (88) |
|   Prefer not to say | 6 (2.8) |
| Occupation | |
|   Student | 37 (17.1) |
|   Health professionals | 13 (6) |
|   Educationalists | 62 (28.5) |
|   Employed other sectors | 90 (41.5) |
|   Own business | 3 (1.4) |
|   Unemployed | 12 (5.5) |
| Location | |
|   Urban | 194 (89.4) |
|   Rural | 23 (10.6) |
| Nationality (most frequent) | |
|   Qatar | 49 (22.6) |
|   Saudi Arabia | 50 (23) |
|   Oman | 31 (14.3) |
|   Jordan | 33 (15.2) |
|   14 Arab countries | 54 (24.9) |
| Frequency of social media usage | |
|   Not frequent | 23 (10.6) |
|   Frequent | 194 (89.4) |
| Frequency of vaccine information search | |
|   Not frequent | 185 (85.2) |
|   Frequent | 32 (14.8) |
| Trust vaccine info shared by contact online | |
|   No trust | 89 (41) |
|   Trust | 128 (58.9) |
| Hearing negative comments about vaccines from social media | |
|   Passive | 79 (36.4) |
|   Active | 138 (63.6) |
| Engage with vaccine discussion in social media | |
|   Passive | 179 (82.5) |
|   Active | 38 (17.5) |

Table 2. Detailed statistics of COVID-19 vaccine acceptance.

| If a COVID-19 vaccine is officially recommended and made available for your age group, profession, and health condition would you be (n = 217 i.e., total number of respondents) | |
|---|---|
| | **Number of respondents** |
| **Total (negative)** | 29 |
|   Very Unlikely to get | 12 |
|   Unlikely to get | 3 |
|   Unsure | 14 |
| **Total (positive)** | 188 |
|   Likely to get | 44 |
|   Very likely to get | 144 |
| When a vaccine for COVID-19 is approved by the government and widely available to anyone who wants it. What will you do? (n = 188). Only applicable for positive responses (i.e., likely to get or very likely to get). | |
| | **Number of respondents** |
| **Total (reluctance)** | 61 |
|   I will wait until it has been available for a while to see how it works for other people. | 42 |
|   I will only get the vaccine if I am required to do it for work, school, or other activities. | 19 |
|   I don't know. | 5 |
| **Total (acceptance)** | 122 |
|   I will get the vaccine as soon as possible. | 116 |
|   Other (Already got vaccine) | 6 |

Table 3. Binary logistic regression to predict vaccine acceptance behavior.

| Social media usage factors | OR | 95% CI | P-Value |
|---|---|---|---|
| Frequency of social media usage | | | |
|   Not Frequent | Ref | | |
|   Frequent | 1.41 | 0.580–3.40 | .45 |
| Frequency of vaccine information search | | | |
|   Not Frequent | Ref | | |
|   Frequent | 1.09 | 0.509–2.35 | .82 |
| Trust vaccine info shared by contacts | | | |
|   Do not trust | Ref | | |
|   Trust | 1.32 | 0.758–2.29 | .328 |
| Hearing negative comments about vaccines | | | |
|   Passive | Ref | | |
|   Active | 0.60 | 0.34–1.08 | .08 |
| Engage with vaccine discussion in social media | | | |
|   Passive | Ref | | |
|   Active | 0.96 | 0.47–1.97 | .915 |

## Public attitudes toward COVID-19 vaccines

Table 4 shows the characteristics of the participant's attitudes toward the COVID-19 vaccines. Some missing values were ignored during the analysis. Of the 217 participants, 51.6% (n = 112) were willing to wait to see the effectiveness of the vaccines, while 57.1% (n = 124) were worried about the side effects of vaccines. However, 53% (n = 115) of the participants were aware that they would not be affected by COVID-19 during vaccination, and 53.9% (n = 117) of them believed that the COVID-19 vaccines would not increase the allergic problems. 25.8% (n = 56) of the participants agreed that vaccines might be harmful to pregnant women. 58.9% (n = 128) of the participants believed that politics played a role in vaccine development. 33.6% (n = 73) of the participants believed that the decision-maker ensured the safety of the vaccine. 47% (n = 102) of the participants assumed that the risk of COVID-19 was being exaggerated. However, 45.2% (n = 98) of the participants responded that they were not free of risk to be infected by the COVID-19. 41.5% (n = 90) of the participants reported that the vaccine development is too fast, and they wanted to wait to see if it works for others. Overall, 54.8% (n = 119) of the participants believed in the efficacy of vaccines.

## Social media usages and public attitudes toward COVID-19 vaccines

Table 5 shows the results obtained from a binary logistic regression that aims to predict the effect of social media usage factors on public attitudes in the COVID-19 vaccine. The table shows only the significant ($p < .05$) factors.

### Social media usage

Risk perception of COVID-19 varies with the usage of social media. Participants who were frequent in social media are 2.853 times more likely to agree that the risk of COVID-19 is being exaggerated (OR = 2.853, 95% CI = 0.862–9.445, $p$ = .046) than those who are not frequent in social media. The confidence interval range 0.862 to 9.445 shows that the data (i.e., frequency of social media usage and risk of COVID-19 being exaggerated) are scattered.

### Trust in vaccine information found online

Individuals are motivated by friends, family, colleagues, and acquaintances. Participants who have more trust in vaccine information shared by their contacts are less likely to agree that decision-makers have verified that vaccines are safe (OR = 0.528, 95% CI = 0.276–1.012, $p$ = .05). Participants who have more trust in the vaccine-related information shared by their contacts online are less likely to be worried about the side effects of vaccines (OR = 0.379, 95% CI = 0.123–1.168, $p$ = .031), are less likely to agree that vaccines may increase allergic problems (OR = 0.393, 95% CI = 0.173–0.893, $p$ = .026), and are likely to trust in vaccines (OR = 0.260, 95% CI = 0.116–0.584, $p$ = .001).

Negative information may affect public attitudes toward COVID-19 vaccines. Active participants (i.e., reply, ask a health professional, ask a friend what they think, search for more information on the internet) are 2.5 times more likely to agree that vaccine might be harmful to pregnant women compared to passive users who ignore the comment (OR = 2.574, 95% CI = 1.080–6.132, $p$ = .033). Also, active participants are more likely to agree that the vaccine development process is fast, and they want to wait to see if vaccines are effective compared to those passive users (OR = 20.019, 95% CI = 0.804–5.069, p = .049).

Table 4. Details of responses related to public attitudes toward COVID-19 vaccines.

| Attitudes | Disagree/Fully disagree, N (Percentages) | Agree/Fully agree, N (Percentages) | Neutral, N (Percentages) | No response, N (Percentages) |
|---|---|---|---|---|
| The vaccine is too new, and I want to wait and see how it works for other people | 41 (18.9) | 112 (51.6) | 63 (29) | 1 (0.4) |
| I am worried about the possible side effects | 32 (14.7) | 124 (57.1) | 49 (22.6) | 12 (5.5) |
| I am worried that I may get COVID-19 from the vaccine | 117 (53.9) | 45 (20.7) | 41 (18.9) | 14 (6.4) |
| I have serious allergic problems and COVID-19 vaccine may increase it further | 115 (53) | 30 (13.8) | 49 (22.6) | 23 (10.6) |
| I think COVID-19 vaccine is harmful for pregnant women | 36 (16.6) | 56 (25.8) | 112 (51.6) | 13 (6) |
| I think politics has played too much of a role in the vaccine development process | 25 (11.5) | 128 (59) | 52 (24) | 12 (5.5) |
| I do not trust the decision makers made sure the vaccine is safe and effective | 73 (33.6) | 78 (36) | 55 (25.3) | 11 (5.1) |
| I think the risks of COVID-19 are being exaggerated | 38 (17.5) | 102 (47) | 64 (29.5) | 13 (6) |
| I do not think I am at risk of getting infection from COVID-19 | 98 (45.2) | 44 (20.2) | 62 (28.6) | 13 (6) |
| I do not trust vaccines in general | 119 (54.8) | 34 (15.6) | 49 (22.6) | 15 (6.9) |
| The vaccine development process is too fast, and I want to wait till the vaccine become mature | 55 (25.3) | 90 (41.5) | 62 (28.6) | 10 (4.6) |

Table 5. Binary logistic regression to predict public attitudes toward the COVID-19 vaccines.

| Factors | OR | 95% CI | P-Value |
|---|---|---|---|
| **Worried about vaccine side effect**[a] | | | |
| Trust vaccine information shared by contacts | | | |
| A little | Ref | | |
| A lot | 0.397 | 0.123–1.168 | .031 |
| **The vaccine may increase the allergic problem** | | | |
| Trust vaccine information shared by contacts | | | |
| A little | Ref | | |
| A lot | 0.393 | 0.173–0.893 | .026 |
| **The vaccine may be harmful to a pregnant woman** | | | |
| Hearing negative comments about vaccines from social media | | | |
| Passive | Ref | | |
| Active | 2.574 | 1.080–6.132 | .033 |
| **I don't trust that decision-makers make sure that vaccines are safe** | | | |
| Trust vaccine info shared by contacts | | | |
| A little | Ref | | |
| A lot | 0.528 | 0.276–1.012 | .05 |
| **Risks of COVID-19 are being exaggerated** | | | |
| Frequency of social media usage | | | |
| Not Frequent | Ref | | |
| Frequent | 2.853 | 0.862–9.445 | .046 |
| **I don't think at risk of getting an infection from COVID-19** | | | |
| Engage with social media discussion | | | |
| Passive | Ref | | |
| Active | 2.636 | 1.077–6.451 | .034 |
| **I don't trust in vaccine** | | | |
| Trust vaccine info shared by contacts | | | |
| A little | Ref | | |
| A lot | 0.260 | 0.116–0.584 | .001 |
| **Vaccine development is too fast and wants to wait and see its effectiveness** | | | |
| Hearing negative comments about vaccines from social media | | | |
| Passive | Ref | | |
| Active | 2.019 | 0.804–5.069 | .049 |

[a]The bold line in the table represents dependent variable. *agree* is coded as 1.

### Engaging in social media discussion

Social media discussions about vaccine-related topics may influence public attitude toward the perceived risk of COVID-19. Active participants (i.e., comment, share, or both) in social media vaccine discussions are 2.64 times more likely to agree that they may not get COVID-19 compared to those who are not active in social media vaccine discussions (OR = 2.636, 95% CI = 1.077–6.451, p = .034).

### Results for multinomial logistic regression analysis

Multinomial logistic regression was applied, and factors (i.e., age, gender, and social media factors) were combined with observing public attitude toward the COVID-19 vaccine. Male participants between 18 and 35 years and who are passive to anti-vaccine comments are less likely to agree that the risk of COVID-19 is being exaggerated (OR = 0.033, 95% CI = 0.001–0.935, p = .046). Male participants of ages between 26 and 35 years and who have no trust in vaccine information shared by contacts are 5.7 times more likely to agree that the risk of COVID-19 is being exaggerated compared to female participants of ages between 18 and 25 and >35 years and who trust vaccine information online (OR = 5.714, 95% CI = 0.001–0.492, P = .017). Male participants of ages more than 35 years and who are active in anti-vaccine comments are 5.417 times more likely to agree that politics played a role in vaccine development (OR = 5.417, 95% CI = 1.055–27.041, p = .043) compared to passive females of ages between 26 and 35 years. Females are less likely to agree that vaccines may increase allergic problems compared to males (OR = 0.404, 95% CI = 0.166–0.983, p = .046).

### Discussion

This study found that social media usage can influence the attitudes of the Arab people toward the COVID-19 vaccines. Social media users connect and exchange ideas on social media platforms, which allows them to reach a large community in a short amount of time. The use of social media is continuously rising among people for sharing health-related information. Consequently, social media users are exposed to health-related information shared by users, including information related to the COVID-19 vaccine too.[19] With greater access to the internet and social media networks, people's opinions regarding vaccination are continuously changing, particularly during the disease outbreaks; vaccine hesitancy has become more dominant among people who frequently search for vaccine-related information on social media.[31] In addition, public confidence in the vaccine may also vary with the collective norms (e.g., social, political, moral, and biological) in different communities.[32] However, false information and fake news regarding vaccines also spread through social media platforms.[33,34] Exposure to misinformation and false claims can affect the decision of people to vaccine acceptance. Also, the activities of influential figures like the health ministry, health professionals, and media figures on social media can affect public attitudes toward vaccines.[35] Similar to our study, another study from KSA was performed to determine the effect of social media on vaccine acceptance.[36] Our study differs in terms of questions and participants' demographics. However, both studies showed that social media might impact public attitudes toward vaccine acceptance.

Our study showed that some of the participants were reluctant to vaccinate. The majority of them were optimistic about vaccine acceptance. However, some participants wanted to delay their vaccination. Vaccine acceptance behavior is often related to the psychological characteristics of human beings and found to be changeable according to the demographic region.[37,38]

The vaccine acceptance predictors are dynamic and related to specific constraints. Participants responded that the COVID-19 vaccines were new, and they wanted to wait until the effectiveness of the COVID-19 vaccines was proved. Some participants expressed that they would take the vaccine only if required for travel, work, and school. The majority of the participants were worried about the possible side effects of vaccines. A good number of participants believed that the risk of COVID-19 is being exaggerated. Some participants were concerned about the role of politics in vaccine development. As the after-effects of COVID-19 were unknown, participants showed their concern about allergic problems, sterility, diabetes, pregnancy, and heart diseases.

Male participants were more willing to vaccinate than female participants. Similar findings were also reported in existing literature.[7–10] Older participants were more likely to get the vaccine than younger participants. Participants who trust vaccine information shared on social media are more concerned about the side effect of the vaccine. It may be because they are exposed to misinformation spread online on social media, as some anti-vaxxers are active on social media to disseminate incorrect information about the side effects of the vaccines.[39] Female participants used social media less than male participants, but they trusted online vaccine information more than males. Female participants are more active in investigating anti-vaccine information, for instance, asking health professionals or friends. On the other hand, male participants engaged more in social media discussion about vaccines compared to female participants. Female participants were more concerned about the side effects of vaccines during pregnancy and breastfeeding and were more likely to be affected more by anti-vaccine contents. Participants were also worried about the long-term side effects of the COVID-19 vaccines as the vaccines were not tested for long-term side effects. However, participants with allergic problems were at risk of side effects more compared to healthy people.

The analysis observed that social media-related variables were not significant factors to predict COVID-19 vaccine intention; however, these variables were significant to predict participant attitudes toward COVID-19 vaccines. It was observed that exposure to vaccine-related information on social media might affect participants' attitudes toward vaccines. For instance, exposure to anti-vaccine contents affected vaccine-specific attitudes such as vaccine may cause side effects, the vaccine may increase allergic problems, and low trust in the decision-maker role to make sure vaccines are safe. Also, influential information (e.g., conspiracy theory, fake news, and misinformation) that prevails in social media might affect people's attitudes, especially female attitudes.[40] Many participants who believed in the information shared by their contacts were concerned about the side effect of the vaccine. Participants who trusted information shared by their contacts were more likely to believe that vaccines may increase the allergic problem. Participants who believed in the vaccine-related information shared by their contacts had trust in the decision-maker role for vaccine safety. Similarly, vaccine-related negative information may also affect participants' attitudes, such as vaccines may be harmful to a pregnant woman and vaccine development process is fast. The effectiveness of vaccines needs to be observed over time. Participants were also concerned that the vaccine might cause infertility. Also, many participants reported concerns about the rapid development of the COVID-19 vaccines, and they wanted to wait to see if vaccines are working for others.

The findings may help the policymakers make effective strategies to overcome the challenges in vaccine acceptance. In addition, the decision-makers can re-organize policies for COVID-19 vaccination by analyzing social media content and raising awareness by sharing reliable news through social media.

## Strength and limitations

### Strength

The study was conducted when COVID-19 vaccines were available, and many people had already received the first dose of the vaccine, so participants had some knowledge of COVID-19 vaccines. The study uses participants' social media usage profiles to find a correlation between public attitudes and the use of social media. The survey was conducted in Arabic and English, so the participants easily understood the questions.

### Limitations

This survey was distributed online, so people who do not use social media and the internet were left out. The survey link was distributed through e-mail, contacts, and social media. No newspapers or television advertised the survey. Hence, the reach to a diversity of people might be limited. The survey could not be performed in offline mode like interviews due to the challenges related to the nationwide lockdown and restrictions on the public gatherings by the time the survey was performed. The number of participants is small, and the results may not reasonably be generalizable for a higher confidence level and a small error margin. This study is based on a convenient self-enrolled sample of respondents that risks not being fully representative of the general population of Arab countries. The sample of respondents was mostly biased with male, young, well-educated, and urban subjects. We did not ask participants whether they or their family members had been infected by COVID-19. This question could have explained more about participants' experiences with COVID-19. During the conduction of the study, people already had some knowledge of the COVID-19 vaccines. The age was not measured on a continuous scale, and age was encoded into groups that may lack some characteristics. The behavior toward vaccines can depend on multiple factors and the use of social media. However, factors other than social media are mostly left out in this survey. The survey was performed from May to June 2021, and public attitudes may have changed. Therefore, further research using other factors is needed to monitor the change in public attitudes.

## Conclusion

While vaccines are useful to contain the spread of the coronavirus, it requires a large portion of the population to be vaccinated. However, vaccine hesitancy is one of the problems that hinder the effort to obtain high vaccination coverage. This study surveyed people in the Arab world to find the impact of social media usage on people's vaccination decisions and attitudes toward COVID-19 vaccines. Although the study did not find any significant correlation between social media usage and people's vaccination decision, there is a notable correlation between social media usage and people's attitude toward COVID-19 vaccines. Social media being a source of public opinion and discussion, bears a significant impact on people's decision-making about health-related issues, especially when access to physical resources is limited and reliance on online information is high during the COVID-19 pandemic. Exposure to COVID-19 vaccine-related misinformation on social media may shape the behavior of the public toward the vaccine. Although public awareness and online campaigns can break the loops of misinformation in social media by promoting vaccine literacy, health-related news should be appropriately framed to effectively communicate the correct information on vaccines. Even though the outcome of this study will be closer to Arab collectivist cultures and social norms, the findings can be relevant internationally.

## Acknowledgment

We acknowledge the contribution of two researchers, Adam G. Dunn and Amalie Dyda for providing useful feedback on the questionnaire. Adam G. Dunn is the director of Biomedical Informatics and Digital Health and Associate Professor in the School of Medical Sciences, Faculty of Medicine and Health, The University of Sydney, Australia. Amalie Dyda is infectious disease epidemiologist working as a teaching and research academic in the School of Public Health at The University of Queensland, Australia.

## Author's contributions

M.R.B., R.A., and Z.S. formulated the research question and developed the questionnaire. M.R.B. and H. A. conducted the survey and performed analysis. M.R.B. and H.A. wrote the initial draft. R.A. and Z.S. revised the draft. All authors agreed on the final results.

## Disclosure statement

No potential conflict of interest was reported by the author(s).

## Funding

The author(s) reported there is no funding associated with the work featured in this article.

## ORCID

Md. Raful Biswas  http://orcid.org/0000-0002-5145-1990
Hazrat Ali  http://orcid.org/0000-0003-3058-5794
Raian Ali  http://orcid.org/0000-0002-5285-7829
Zubair Shah  http://orcid.org/0000-0001-7389-3274

# Appendix A

**Questionnaires**
**Covid-19 vaccine perception**
*Section A*
*General Information*

| 1 | Age | ____________ |
|---|---|---|
| 2 | Gender | ▫ Male |
|   |   | ▫ Female |
|   |   | ▫ Prefer not to say |
| 3 | I live with | ▫ Family |
|   |   | ▫ Alone |
|   |   | ▫ Other People |
|   |   | ▫ Prefer not to say |
| 4 | Occupation | ▫ Health Professionals |
|   |   | ▫ Educationalists |
|   |   | ▫ Employed—Other Sectors |
|   |   | ▫ Own Business |
|   |   | ▫ Unemployed |
|   |   | ▫ Student |
| 5 | Nationality | ▫ Qatar |
|   |   | ▫ Saudi Arabia |
|   |   | ▫ United Arab Emirates |
|   |   | ▫ Kuwait |
|   |   | ▫ Bahrain |
|   |   | ▫ Oman |
|   |   | ▫ Iraq |
|   |   | ▫ Syria |
|   |   | ▫ Palestinian |
|   |   | ▫ Jordan |
|   |   | ▫ Lebanon |
|   |   | ▫ Libya |
|   |   | ▫ Other.____________ |
| 6 | Country of residence | ▫ Qatar |
|   |   | ▫ Saudi Arabia |
|   |   | ▫ United Arab Emirates |
|   |   | ▫ Kuwait |
|   |   | ▫ Bahrain |
|   |   | ▫ Oman |
|   |   | ▫ Iraq |
|   |   | ▫ Syria |
|   |   | ▫ Palestinian |
|   |   | ▫ Jordan |
|   |   | ▫ Lebanon |
|   |   | ▫ Libya |
|   |   | ▫ Other.____________ |
| 7 | Location | ▫ Urban area |
|   |   | ▫ Rural area |

**Section B**
*Vaccine Specific Question*

| | | |
|---|---|---|
| 8 | How often do you use social media? | ▫ Never<br>▫ Once a month or less<br>▫ Several times a month<br>▫ Once a week<br>▫ More than once a week<br>▫ Once a day<br>▫ More than once a day |
| 9 | How often do you search information about COVID-19 vaccine? | ▫ Never<br>▫ Once a month or less<br>▫ Several times a month<br>▫ Once a week<br>▫ More than once a week<br>▫ Once a day<br>▫ More than once a day |
| 10 | How much do you trust information about COVID-19 vaccines shared by your contacts online? | ▫ A great deal<br>▫ A lot<br>▫ A moderate amount<br>▫ A little<br>▫ Not at all |
| 11 | If you hear a negative comment about COVID-19 vaccine on social media, what would you do? | ▫ I ignore it<br>▫ I reply to it<br>▫ I ask friends what they think<br>▫ I ask a health professional about it<br>▫ I search for more information on the internet |
| 12 | How much do you engage with COVID-19 vaccine discussion on social media? | ▫ I just read it<br>▫ I comment on it<br>▫ I share it<br>▫ I share it and comment on it<br>▫ I ignore it |
| 13 | Did you get the COVID-19 vaccination? | ▫ Yes<br>▫ No |
| 14 | If a COVID-19 vaccine is officially recommended and made available for your age group, profession and heath condition you would be | ▫ Very likely to get it    ▫ Unsure<br>▫ Likely to get it    ▫ Unlikely to get it<br>    ▫ Very unlikely to get it<br>Please go to Question 15    Please go to Question 16 |
| 15 | When a vaccine for COVID-19 is approved by the government and widely available to anyone who wants it. What will you do? | ▫ I will get the vaccine as soon as possible.<br>▫ I will wait until it has been available for a while to see how it is working for other people<br>▫ I will only get the vaccine if I am required to do so for work, school, or other activities<br>▫ I don't know<br>▫ Other |

16 How much do you agree with the following statements?

| | Full agree | Agree | Neutral | Full Disagree | Disagree |
|---|---|---|---|---|---|
| The vaccine is too new, and I want to wait and see how it works for other people | ▫ | ▫ | ▫ | ▫ | ▫ |
| I am worried about the possible side effects | ▫ | ▫ | ▫ | ▫ | ▫ |
| I am worried that I may get COVID-19 from the vaccine | ▫ | ▫ | ▫ | ▫ | ▫ |
| I have serious allergic problems and COVID-19 vaccine may increase it further | ▫ | ▫ | ▫ | ▫ | ▫ |
| I think COVID-19 vaccine is harmful for pregnant women | ▫ | ▫ | ▫ | ▫ | ▫ |
| I think politics has played too much of a role in the vaccine development process | ▫ | ▫ | ▫ | ▫ | ▫ |
| I do not trust the decision makers made sure the vaccine is safe and effective | ▫ | ▫ | ▫ | ▫ | ▫ |
| I think the risks of COVID-19 are being exaggerated | ▫ | ▫ | ▫ | ▫ | ▫ |
| I do not think I am at risk of getting infection from COVID-19 | ▫ | ▫ | ▫ | ▫ | ▫ |
| I do not trust vaccines in general | ▫ | ▫ | ▫ | ▫ | ▫ |
| The vaccine development process is too fast and I want to wait till the vaccine become mature | ▫ | ▫ | ▫ | ▫ | ▫ |

17 If you have any other concerns about COVID-19 vaccines, please type them here  ____________________
18 If you know of other people concerns about COVID-19 vaccines, please type them here  ____________________

*(Continued)*

(Continued).

| 19 | Do you have a Twitter account? We will do automated analysis to match the responses of this survey to information people communicate on Twitter. Your identity will remain strictly confidential. We will not look to identify you and will delete all the data after we are done with the analysis. Your identification will never appear in any published work. If you agree and have Twitter, please provide your Twitter handle here: | __________ |
|---|---|---|

# Appendix B

**Table B1.** Data encoding table.

| Age-band | Category | Encoding |
|---|---|---|
| | 18-25 | 1 |
| | 26–35 | 2 |
| | >35 = 3 | 3 |
| **Gender** | | |
| | Male | 1 |
| | Female | 2 |
| | Prefer not to say | 3 |
| **Live with** | | |
| | Alone | 1 |
| | Family | 2 |
| | Prefer not to say | 3 |
| **Occupation** | | |
| | Student | 1 |
| | Health Professionals | 2 |
| | Educationalists | 3 |
| | Employed other sectors | 4 |
| | Own Business | 5 |
| | Unemployed | 6 |
| **Place of living** | | |
| | Urban | 1 |
| | Rural | 2 |
| **How often do you use social media?** | | |
| **How often do you search for information about the COVID-19 vaccine?** | | |
| Never | Not frequent | 1 |
| Once a month or less | | |
| Several times a month | | |
| Once a week | | |
| More than once a week | | |
| More than Once a day | Frequent | 2 |
| Once a day | | |
| **How much do you trust information about COVID-19 vaccines shared by your contacts online?** | | |
| Not at all | Do not trust | 1 |
| A little | | |
| A moderate amount | Trust | 2 |
| A lot | | |
| A great deal | | |
| **If you hear a negative comment about the COVID-19 vaccine on social media, what would you do?** | | |
| Ignore it | Passive | 1 |
| Reply to it | Active | 2 |
| Ask a health professional about it | | |
| Ask a friend what they think about it | | |
| Search for more information on the internet | | |
| **How do you engage with COVID-19 vaccine discussion on social media?** | | |
| Just read it | Passive | 1 |
| Ignore it | | |

(*Continued*)

**Table B1.** (Continued).

| Age-band | Category | Encoding |
|---|---|---|
| Comment on it | Active | 2 |
| Share it | | |
| Share it and comment on it | | |
| **If a COVID-19 vaccine is officially recommended and made available for your age group, profession and health condition would you be** | | |
| Very unlikely to get it | Hesitance | 1 |
| Unlikely to get it | | |
| Unsure | | |
| Likely to get it | Acceptance | 3 |
| Very likely to get it | | |
| **When a vaccine for COVID-19 is approved by the government and widely available to anyone who wants it. What will you do?** | | |
| I will wait until it has been available for a while to see how it is working for other people | Hesitant | 1 |
| I will only get the vaccine if I am required to do so for work, school, or other activities | | |
| I will get the vaccine as soon as possible | Acceptance | 2 |
| **Vaccine acceptance (Custom variable)** | | |
| Very unlikely to get it | Hesitant | 1 |
| Unlikely to get it | | |
| Unsure | | |
| I will wait until it has been available for a while to see how it is working for other people | | |
| I will only get the vaccine if I am required to do so for work, school, or other activities | | |
| I will get the vaccine as soon as possible | Acceptance | 2 |
| **Likert scale** | | |
| Fully Disagree, Disagree | Disagree | 1 |
| Neutral | Neutral | 2 |
| Agree, Fully Agree | Agree | 3 |